\documentclass{PoS}
\usepackage{amsmath}
\usepackage{amssymb}
\usepackage{bm}
\usepackage[dvips]{epsfig}
\usepackage{color}
\usepackage{times}
\usepackage{subfigure}
\usepackage{cite}
\usepackage{epstopdf}

\newcommand{\Dmesonneut}{\overline{D^{0}}}

\newcommand{\mylambda}{\Lambda_{c}^{+}}

\newcommand{\process}{ p ~ \gamma \rightarrow \mylambda ~\Dmesonneut}

\title{Hard exclusive photoproduction of charmed mesons}

\ShortTitle{$ \process $ photoproduction}

\author{Alexander Thomas Goritschnig \\
        Centre de Physique Th\' eorique, \' Ecole Polytechnique, 91128 Palaiseau cedex, France \\
        Institute of Physics, University of Graz, 8010 Graz, Austria \\
        E-mail: \email{alexander.goritschnig@uni-graz.at}}

\author{\speaker{Stefan Kofler} \\
        Institute of Physics, University of Graz, 8010 Graz, Austria \\
        E-mail: \email{stefan.kofler@uni-graz.at}}

\author{Wolfgang Schweiger \\
        Institute of Physics, University of Graz, 8010 Graz, Austria\\
        E-mail: \email{wolfgang.schweiger@uni-graz.at}}

\abstract{We investigate the photoproduction process $\process$ within the handbag approach,
which we assume to be the dominant mechanism at energies well above the production threshold and in the forward scattering hemisphere.}

\FullConference{Photon 2013,\\
		20-24 May 2013\\
		Paris, France }

\begin{document}

\section{Introduction}
\label{sec:Introduction}
The purpose of this contribution is to study the exclusive photoproduction of charmed mesons, or to be more specific the reaction\footnote{Among the reactions \( \gamma + N \rightarrow B_c + \overline{D} \) with \(B_c = \Lambda_c \) or \( \Sigma_c \), the reaction \( \process \) has the lowest threshold.} \( \process \).
Until now only inclusive measurements of open charm production exist, no exclusive ones.
This could change with the planned \( 12~\mbox{GeV} \) upgrade at JLAB. Therefore it is interesting to estimate the cross section for that process.
We  investigate it within the generalized parton picture assuming that the, so-called, \lq\lq handbag mechanism\rq\rq\ dominates this process well above the production threshold.
We argue that under physical plausible assumptions the photoproduction amplitude factorizes into a perturbatively calculable partonic subprocess and hadronic matrix elements, which contain the non-perturbative bound-state dynamics.
 These hadronic matrix elements are  parameterized in terms of generalized parton distributions and a meson distribution amplitude.
With simple models for the hadron light-cone wave functions, which give us the generalized parton distributions for the $p\rightarrow \Lambda_c$ transition and the \( D \)-meson distribution amplitude, we obtain numerical predictions for $p\rightarrow \mylambda$ transition form factors and for the differential and integrated \( \process \) cross sections.

\section{Kinematics}
\label{sec:Kinematics}
The momenta of the incoming proton and photon are denoted by \(p \) and \(q\), those of the outgoing \(\mylambda\) and \(\Dmesonneut\) by \( p^\prime\) and \( q^\prime \), respectively.
We work in a center-of-momentum system (CMS) which has its \( z \)-axis aligned along \( \bar{\mathbf{p}} \), the spatial part of the average momentum \( \bar{p} \equiv \frac{1}{2} \left(p + p^{\prime}\right)\). In this frame the transverse component of the momentum transfer \( \Delta \equiv \left(p^\prime- p  \right) = \left( q - q^\prime \right) \) is symmetrically shared between the particles.
The momenta of the particles are parameterized as
\begin{equation}
\begin{split}
& p =        \left[ (1 + \xi) \bar{p}^{+},\, \frac{m^{2} + {\bm \Delta}_\perp^2/4}{2(1 + \xi) \bar{p}^{+}},\, -\frac{{\bm \Delta}_\perp}{2} \right]\,, \quad
  p^\prime = \left[ (1 - \xi) \bar{p}^{+},\, \frac{M^{2} + {\bm \Delta}_\perp^2/4}{2(1 - \xi) \bar{p}^{+}},\, +\frac{{\bm \Delta}_\perp}{2} \right]\,, \\
& q =        \left[ \frac{{\bm \Delta}_\perp^2/4}{2(1 + \eta) \bar{q}^{-}},\, (1 + \eta) \bar{q}^-,\, +\frac{{\bm \Delta}_\perp}{2} \right]\,, \quad
  q^\prime = \left[ \frac{M_D^2 + {\bm \Delta}_\perp^2/4}{2(1 - \eta) \bar{q}^{-}},\, (1 - \eta) \bar{q}^-,\, -\frac{{\bm \Delta}_\perp}{2} \right]\,,
\label{eq:momenta}
\end{split}
\end{equation}
where we have introduced a second average momentum
\begin{equation}
 \bar{q} \,:=\, \frac12 \left( q^\prime + q \right)
\label{eq:pbar-qbar-def}
\end{equation}
and the skewness parameters
\begin{equation}
 \xi \,:=\, - \frac{\Delta^+}{2\bar{p}^+}
\quad\text{and}\quad
 \eta \,:=\,  \frac{\Delta^-}{2\bar{q}^+}\, .
\label{eq:def-skewness-parameters}
\end{equation}
It should be noted that $\bar{q}$ and $\eta$ are not independent, but rather functions of $\bar{p}$ and $\xi$. They have just been introduced for brevity of notation and to make our analytical formulae better readable.

\begin{figure}[h]
\centering
\includegraphics[width=.6\textwidth]{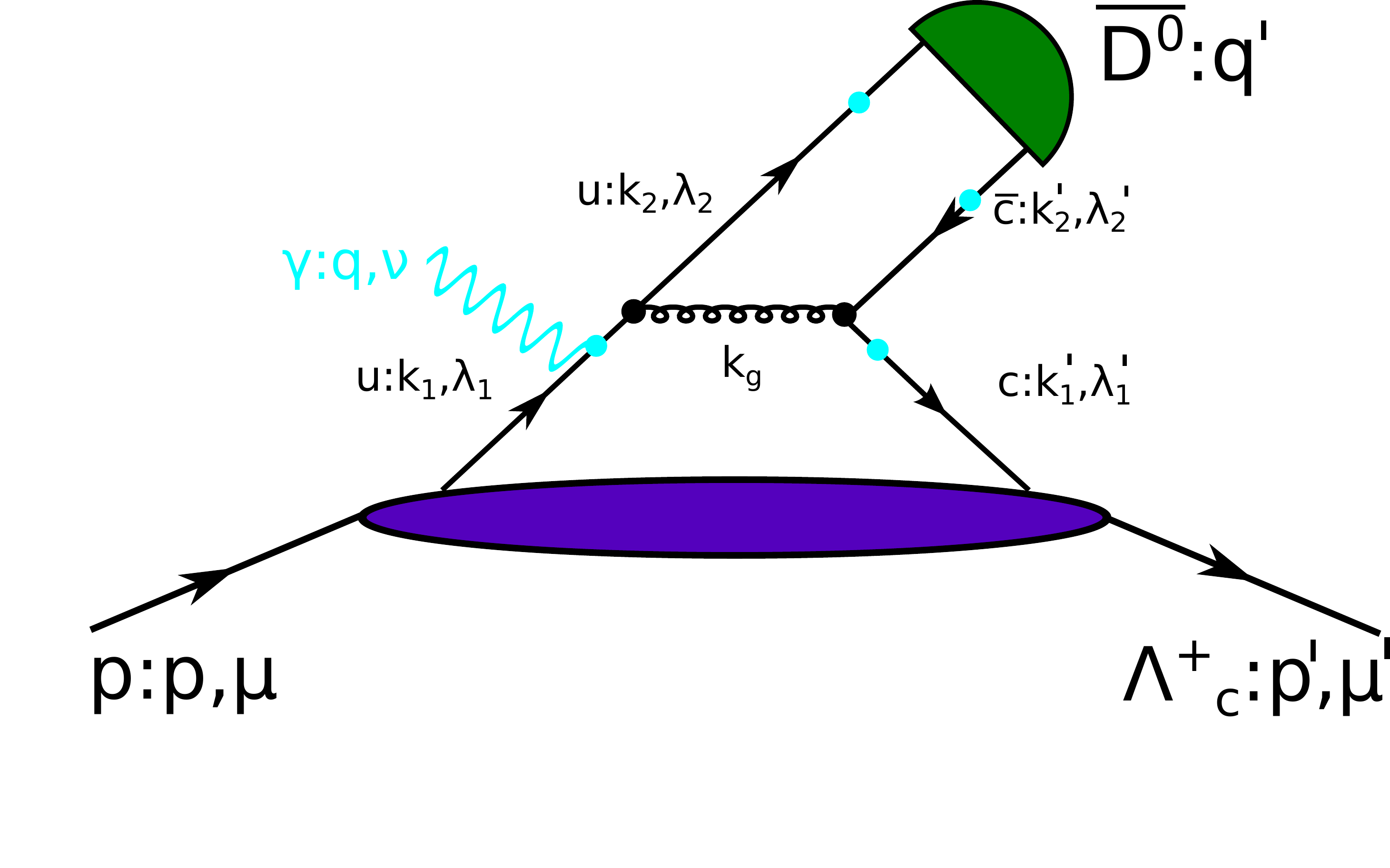}
\caption{The handbag contribution to the proces \(\process \) (in the DGLAP region). The momenta and helicities of the hadrons and partons are specified. The charm content of the proton is is assumed to be negligible. The photon can couple to either of the incoming or outgoing quark lines, as indicated by the dots.}
\label{fig_handbag}
\end{figure}

\section{The Handbag Mechanism}
\label{sec:The-Handbag-Mechansim}
We assume that a handbag-type mechanism dominates the photoproduction process $ \process$ at CMS energies well above the production threshold and in the forward scattering hemisphere.
Its schematics is depicted in Fig.~\ref{fig_handbag}.
In the handbag approach the $ \process$ process amplitude factorizes
into a hard subprocess amplitude on the parton level and
into hadronic matrix elements describing the soft hadronic transitions on the hadron level.

Characteristic for this mechanism is that only the minimal number of hadronic constituents, which is necessary to convert the incoming particles into the final ones, actively takes part in the subprocess .
Since we neglect intrinsic proton charm it is the subprocess $\gamma u \to c\bar{c} u$ which we have to consider.
In order to produce the heavy $c$-quark pair in the subprocess the gluon has to be highly virtual.
To justify a perturbative QCD treatment of the subprocess also the quark propagators between the photon and gluon vertex have to be highly virtual.
That this is the case can easily be checked numerically~\cite{Kofler:2013}.
This justifies to call the subprocess a hard one and to treat it by means of perturbative QCD.

On the other hand one has Fourier transforms of soft hadronic matrix elements.  These describe the non-perturbative
formation of the pseudoscalar $\overline{D^0}$ meson out of the $u$ and $\bar{c}$ quark coming from the hard subprocess
and the non-perturbative transition from the proton to the $\Lambda_c^+$ via the emission of a $u$ quark, that subsequently enters the hard subprocess and the absorption of the $c$ quark which comes lout of the hard subprocess.
A bilocal product of a $u$- and a $c$-quark operator is sandwiched between the vacuum and the $\overline{D^0}$ state for the description of the $\overline{D^0}$ formation,
whereas it is sandwiched between the proton and the $\Lambda_c^+$ state for the description of the $p \to \Lambda_c^+$ transition.

In order to justify factorization we have posed physically motivated restrictions on the virtualities and intrinsic transverse momenta of the partons that parallel the ones of Refs.~\cite{Diehl:1998, Huang:2000, schweig1, schweig2}.
As one consequence the active parton momenta are approximately on mass shell and collinear with their parent hadron momenta
\begin{equation}
\begin{split}
 & k_{1}          \approx  \left[ x_1 (1 + \xi) \bar{p}^{+},\, \frac{x_{1}^2 {\bm\Delta}_{\perp}^{2}/4}{2 x_1 (1 + \xi) \bar{p}^{+}},\, - x_1 \frac{{\bm\Delta}_{\perp}}{2} \right] \,,\quad
   k_{1}^{\prime}         \approx  \left[ x_1^{\prime} (1 - \xi) \bar{p}^{+},\, \frac{m_{c}^{2} + x_{1}^{\prime 2} {\bm\Delta}_{\perp}^{2}/4}{2 x_1^{\prime} (1 - \xi) \bar{p}^{+}},\, + x_1^{\prime} \frac{{\bm\Delta}_{\perp}}{2} \right]\,, \\
 & k_{2} \approx  \left[ \frac{y_1^2 {\bm\Delta}_{\perp}^{2}/4}{2y_1 (1 - \eta) \bar{q}^-},\, y_1 (1 - \eta) \bar{q}^-,\, - y_1 \frac{{\bm\Delta}_{\perp}}{2} \right] \,,\quad
   k_{2}^{\prime} \approx  \left[ \frac{m_{c}^{2} + y_2^2 {\bm\Delta}_{\perp}^{2}/4}{2 y_2 (1 - \eta) \bar{q}^-},\, y_2 (1 - \eta) \bar{q}^-,\, - y_2 \frac{{\bm\Delta}_{\perp}}{2} \right] \,,
\label{eq:parton-momenta}
\end{split}
\end{equation}
where $ x_1 := k_{1}^{+}/p^{+} $, $ x_1^{\prime} := k_{1}^{\prime +}/p^{ \prime +} $ and $ y_1 = 1 - y_2 $ with $ y_2 := k_{2}^{\prime -}/q^{\prime -} $.
For later purposes we also introduce the average momentum fraction $\bar{x} := \bar{k}^+ / \bar{p}^+$ where $\bar{k} := (k_1 + k_1^\prime)/2$.
It can be related to $x_1$ and $x_1^{\prime}$ by
\begin{equation}
x_1 \,=\, \frac{\bar{x}+\xi}{1+\xi}
\quad\text{and}\quad
x_1^{\prime} \,=\, \frac{\bar{x}-\xi}{1-\xi} \,,
\label{eq:x1-and-x2-wrt-xbar}
\end{equation}
respectively.
Another consequence is that the relative distances between the quark-field operators in the hadronic matrix element
for the $p \to \Lambda_c^+$ transition and the $\overline{D^0}$ formation are forced to be lightlike, i.e. the operators act at the same light-cone (LC) time and thus there is no further need for time ordering of the  quark fields in the hadronic matrix elements.
Hence the $\process$ amplitude can be written as
\begin{equation}
\begin{split}
 \mathcal{M}_{\mu^\prime0, \mu \nu} \,=\,
 &\int   dk_2^{\prime -} \, \theta(k_2^{\prime -}) \, \int   d\bar{x}\, \bar{p}^+ \,
  \int \, \frac{dz_1^-}{2\pi} \, e^{\imath\bar{x}\bar{p}^+z_1^-} \,
   \langle \, \Lambda_c^+ :\, p^\prime,\,\mu^\prime \, \mid
    \overline{\Psi}^c \left(-\frac{z_1^-}{2}\right) \Psi^u \left(+\frac{z_1^-}{2}\right)
   \mid \, p:\, p,\,\mu \, \rangle \\
 &\times \,\, \Tilde{H}^\nu \left( \bar{x},\,k_2^{\prime -},\, q^\prime \right) \,
  \int \, \frac{dz_2^+}{2\pi} \, e^{\imath k_2^{\prime -} z_2^+} \,
   \langle \, \overline{D^0} :\, q^\prime \, \mid
    \overline{\Psi}^u \left(z_2^+ \right) \Psi^c(0)
   \mid \, 0 \rangle \,,
\label{eq:process-amplitude}
\end{split}
\end{equation}
where color and spinor labels have been suppressed for the ease of writing.
$\Tilde{H}^\nu$ denotes the hard scattering kernel of the subprocess,
the first and second Fourier transform comprise the soft transition from the proton to the $\Lambda_c^+$
and the $\overline{D^0}$ formation, respectively.
\section{Hadronic Matrix Elements and the Double Handbag Amplitude}
\label{sec:The-Hadronic-Matrix-Elements}
For the non-perturbative $p \to \Lambda_c^+$ transition and the $\overline{D^0}$ formation one now has to study
\begin{equation}
  \bar{p}^+ \, \int \, \frac{dz_1^-}{2\pi} \, e^{\imath\bar{x}\bar{p}^+z_1^-} \,
  \langle \, \Lambda_c^+ :\, p^\prime,\,\mu^\prime \, \mid
   \overline{\Psi}^c(-\frac{z_1^-}{2}) \Psi^u(+\frac{z_1^-}{2})
  \mid \, p:\, p,\,\mu \, \rangle \,,
\label{eq:p-to-lambdac-transition}
\end{equation}
and
\begin{equation}
  \int \, \frac{dz_2^+}{2\pi} \, e^{\imath k_2^{\prime -} z_2^+} \,
   \langle \, \overline{D^0} :\, q^\prime \, \mid
    \overline{\Psi}^u(z_2^+) \Psi^c(0)
   \mid \, 0 \rangle \,,
\label{eq:D0-formation}
\end{equation}
respectively.

In Ref.~\cite{schweig1} the $p \to \Lambda_c^+$ matrix element~(\ref{eq:p-to-lambdac-transition}) has already been analyzed in connection with the process $p\bar{p} \,\to\, \Lambda_c^+\overline{\Lambda}_c^-$.
There it has been shown that it can be decomposed into other matrix elements
\begin{equation}
 \sum_{\lambda_1} \,
 \Big\{
  \left[ \mathcal{H}_{\mu^\prime\mu}^{cu\,+} + 2\lambda_1 \mathcal{H}_{\mu^\prime\mu}^{cu\,+5} \right]
     \frac{\bar{u}(k_2,\,\lambda_1) u(k_1,\,\lambda_1)}{4\sqrt{k_1^+k_2^+}}
  -\lambda_1 \left[ \mathcal{H}_{\mu^\prime\mu}^{cu\,+1} - 2\lambda_1 \imath \mathcal{H}_{\mu^\prime\mu}^{cu\,+2} \right]
     \frac{\bar{u}(k_2,\,-\lambda_1) u(k_1,\,\lambda_1)}{4\sqrt{k_1^+k_2^+}}
 \Big\}
\label{eq:p-to-lambdac-transition-decomposed}
\end{equation}
which are associated with various Dirac structures with the dominant contributions coming from those with $\gamma^+,\gamma^+ \gamma_5$ and \(\sigma^{+j} = \imath \gamma^+ \gamma^j\) ($j = 1,\,2$):
\begin{equation}
 \mathcal{H}_{\mu^\prime\mu}^{cu\,\{+,\, +5,\, +j\}} \,:=\,
  \bar{p}^+ \int \frac{dz_1^-}{2\pi} e^{\imath\bar{x}\bar{p}^+z_1^-}
  \langle \, \Lambda_c^+ :\, p^\prime,\,\mu^\prime \, \mid
  \overline{\Psi}^c(-\frac{z_1^-}{2}) \{ \gamma^+,\, \gamma^+\gamma_5,\, \imath\sigma^{+j} \} \Psi^u(+\frac{z_1^-}{2})
  \mid \, p :\, p,\,\mu \, \rangle \, .
\label{eq:def-mathcal-Hs}
\end{equation}
Each hadronic matrix element in Eq.~(\ref{eq:def-mathcal-Hs}) is then expanded in terms of Lorentz covariants, which depend only on hadronic momenta and spins: 
\begin{eqnarray}
 \mathcal{H}_{\mu^\prime\mu}^{cu\,+} &=&
  \bar{u}(p^\prime , \mu^\prime) \left[
   H^{cu}(\bar{x},\xi,t) \, \gamma^+
   \,+\,
   E^{cu}(\bar{x},\xi,t)\,\frac{\imath\sigma^{+\nu} \Delta_\nu}{M+m}
  \right] u(p , \mu),
\label{eq:mathcalH} \\
 \mathcal{H}_{\mu^\prime\mu}^{cu\,+5} &=&
  \bar{u}(p^\prime , \mu^\prime) \left[
   \widetilde{H}^{cu}(\bar{x},\xi,t) \, \gamma^+\gamma_5
   \,+\,
   \widetilde{E}^{cu}(\bar{x},\xi,t) \, \frac{\Delta^+}{M+m}\gamma_5
  \right] u(p ,\mu),
\label{eq:mathcalHTilde} \\
 \mathcal{H}_{\mu^\prime\mu}^{cu\,+j} &=&
  \bar{u}(p^\prime , \mu^\prime) \left[
    H^{cu}_T(\bar{x},\xi,t) \, \imath \sigma^{+j}
    +
    \widetilde{H}^{cu}_T(\bar{x},\xi,t) \, \frac{\bar{p}^{+}\Delta^j-\Delta^+\bar{p}^j}{Mm} \right. \nonumber \\
    &\phantom{=}& \left. + E^{cu}_T(\bar{x},\xi,t) \, \frac{\gamma^+\Delta^j-\Delta^+\gamma^j}{M+m}
    + \widetilde{E}^{cu}_T(\bar{x},\xi,t) \, \frac{\gamma^+\bar{p}^j-\bar{p}^{+}\gamma^j}{(M+m)/2}
  \right] u(p , \mu).
\label{eq:mathcalHT}
\end{eqnarray}
The coefficients in front of these covariants are the $p \to \Lambda_c^+$ generalized parton distributions (GPDs) which are off-diagonal in flavor space. They depend on the average momentum fraction $\bar{x}$, the skewness parameter $\xi$ and Mandelstam $t = \Delta^2$.
A closer inspection of Eqs.~(\ref{eq:def-mathcal-Hs})-(\ref{eq:mathcalHT}) reveals that
only the GPDs $H^{cu}$, $\widetilde{H}^{cu}$ and $H^{cu}_T$
do not have the need for orbital angular momenta of the partons.
Assuming the ground-state wave function of proton and $\Lambda_c$ to be predominantly an s wave
therefore means that the $H^{cu}$, $\widetilde{H}^{cu}$ and $H^{cu}_T$ contributions should be the most important ones. Thus we will neglect in the following all other contributions except those coming with
$H^{cu}$, $\widetilde{H}^{cu}$ and $H^{cu}_T$.

The hadronic matrix element (\ref{eq:D0-formation}) for the $\overline{D^0}$ formation can be written as~\cite{Huang:2000}
\begin{equation}
\frac{f_D}{2\sqrt{6}} \, \frac{ ( \gamma^\mu q^\prime_\mu - M_D )\gamma_5 }{ \sqrt{2} } \, \phi_D(y_1) \,,
\label{eq:D0-formation-parametrization}
\end{equation}
where $f_D$ is the decay constant and $\phi_D$ the DA for the pseudoscalar $D^0$ meson.
We now have all ingredients to write down the \( \process\) photoproduction amplitude.
But before doing so we make a further simplification by applying the, so-called, ``peaking approximation''.
It was argued in Ref.~\cite{schweig1} that, due the the heavy \(m_c\) mass, the \(p \rightarrow \mylambda \) GPDs are peaked at \( x_0 ={m_c}/{M} \).
This allows us to replace the hard-scattering part of the process amplitude by its value at \( x_0 \).
Thus, the hard subprocess amplitude can be taken out from the $\bar{x}$ integral, which then only has to be performed over the GPDs.
With the peaking approximation our LC-helicity amplitudes for the $\process$ photoproduction finally become ($\nu$ is the LC-helicity of the photon):
\begin{equation}
 \mathcal{M}_{\pm\pm}^\nu =
  \frac{\sqrt{1-\xi^2} \, f_D}{9\sqrt{2}} \, \left[ \left( R_V \pm R_A \right) H_{++}^\nu + \left( R_V \mp R_A \right) H_{+-}^\nu \right] \,,\quad
 \mathcal{M}_{\pm\mp}^\nu  =
  \sqrt{1-\xi^2} \, \frac{2f_D}{9\sqrt{2}} \, S_T \, H_{\pm\mp}^\nu \,,
\label{eq:helicity-amplitudes}
\end{equation}
where we have introduced
\begin{equation}
 H_{\lambda_2\lambda_1}^\nu \,:=\,
  \int_0^1 dy_1 \phi_D(y_1) \, \bar{u}(k_1^\prime,\,\lambda_1^\prime) \Tilde{H}^\nu\left( \bar{x},\, y_1,\, q \right) u(k_1,\,\lambda_1)
 \quad\text{with}\quad
  \bar{x} \,\to\, x_0
 \label{eq:subprocess-amplitude}
\end{equation}
and the generalized $p \to \Lambda_c^+$ form factors
\begin{equation}
 \{ R_V,\, R_A,\, S_T \} \,:=\, \int_\xi^1 \, \frac{d\bar{x}}{\sqrt{\bar{x}^2-\xi^2}} \, \{ H^{cu},\, \widetilde{H}^{cu}\,, H^{cu}_T \} \,.
\label{eq:def-form-factors}
\end{equation}

\section{Modeling the Hadronic Transitions}
\label{sec:Modelling-the-Hadronic-Transitions}
In order to make numerical predictions one has to specify the $p\,\to\Lambda_c^+$ GPDs.
We follow Ref.~\cite{Diehl:2000xz} and model the GPDs by means of an overlap of pertinent light-cone wave functions (LCWFs). Within this construction one makes use of the Fourier expansion of the quark-field operators and the Fock-state decomposition of the hadron states in the LC formalism.
For our purposes we only take the valence Fock states of the proton and the $\Lambda_c^+$ into account
and consider only contributions with zero orbital angular momenta of the partons.
We furthermore assume that the heavy $c$ quark carries the same helicity as its parent $\Lambda_c^+$.
\footnote{An opposite helicity contribution has been considered Ref.~\cite{schweig1}, but turned out to be almost negligible.}
Thus we only need one LCWF for the proton and one for the $\Lambda_c^+$ for which we take as in Ref.~\cite{schweig1}
\begin{equation}
 \psi_p = N_p \, ( 1 + 3x_1 ) \, e^{- a_p^2 \sum_{i=1}^3 \frac{{\bm k}_{i\perp}^2}{x_i}}
  \quad\text{and}\quad
 \psi_\Lambda = N_\Lambda \, e^{- a_\Lambda^2 M^2 \frac{(x_1-x_0)^2}{x_1(1-x_1)}} \, e^{- a_\Lambda^2 \sum_{i=1}^3 \frac{{\bm k}_{i\perp}^2}{x_i}} \,,
\label{eq:p-lambdac-LCWFs}
\end{equation}
respectively.
The proton LCWF is a well established one and was first proposed in Ref.~\cite{Bolz:1996sw}.
The LCWF for the $\Lambda_c^+$ was introduced in Ref.~\cite{schweig1}. The additional $x$-dependent exponential was suggested in Ref.~\cite{Korner:1992uw}. It provides a maximum at $x_0$ in accordance with expectations from heavy-quark effective theory~\cite{Isgur}.
These simple wave function models contain four open parameters, the transverse size parameters $a_{p/\Lambda}$ and the normalization constants $N_{p/\Lambda}$.
We choose, as in Refs.~\cite{Bolz:1996sw, schweig1}, $a_p=0.75~\text{GeV}^{-1}$, $N_p=160.93~\text{GeV}^{-2}$, 
$a_\Lambda=0.75~\text{GeV}^{-1}$ and $N_\Lambda=3477~\text{GeV}^{-2}$.

The non-perturbative input for the $\overline{D^0}$ formation, the $\overline{D^0}$ DA,
can also be derived from a corresponding LCWF by integrating it over ${\mathbf k}_\perp$.
We take, in analogy to the $\Lambda_c^+$ LCWF~(\ref{eq:p-lambdac-LCWFs}),
\begin{equation}
 \psi_D = N_D \, e^{- a_D^2 M_D^2 \frac{(y_1-m_c/M_D)^2}{y_1(1-y_1)}} \, e^{- a_D^2 \frac{{\bm k}_{\perp}^2}{y_1(1-y_1)}} \,,
\label{eq:D-LCWF}
\end{equation}
where we fix $a_D=0.86~\text{GeV}^{-1}$ and $N_D=55.19~\text{GeV}^{-2}$
such that a valence-Fock-state probability of $0.9$ and $f_D = 206~\text{MeV}$~\cite{PDG} (meson decay constant) are obtained.

\section{Cross Section Predictions}
\label{sec:Cross-Section-Predictions}

\begin{figure}[h]
\centering
\includegraphics[width=.45\textwidth]{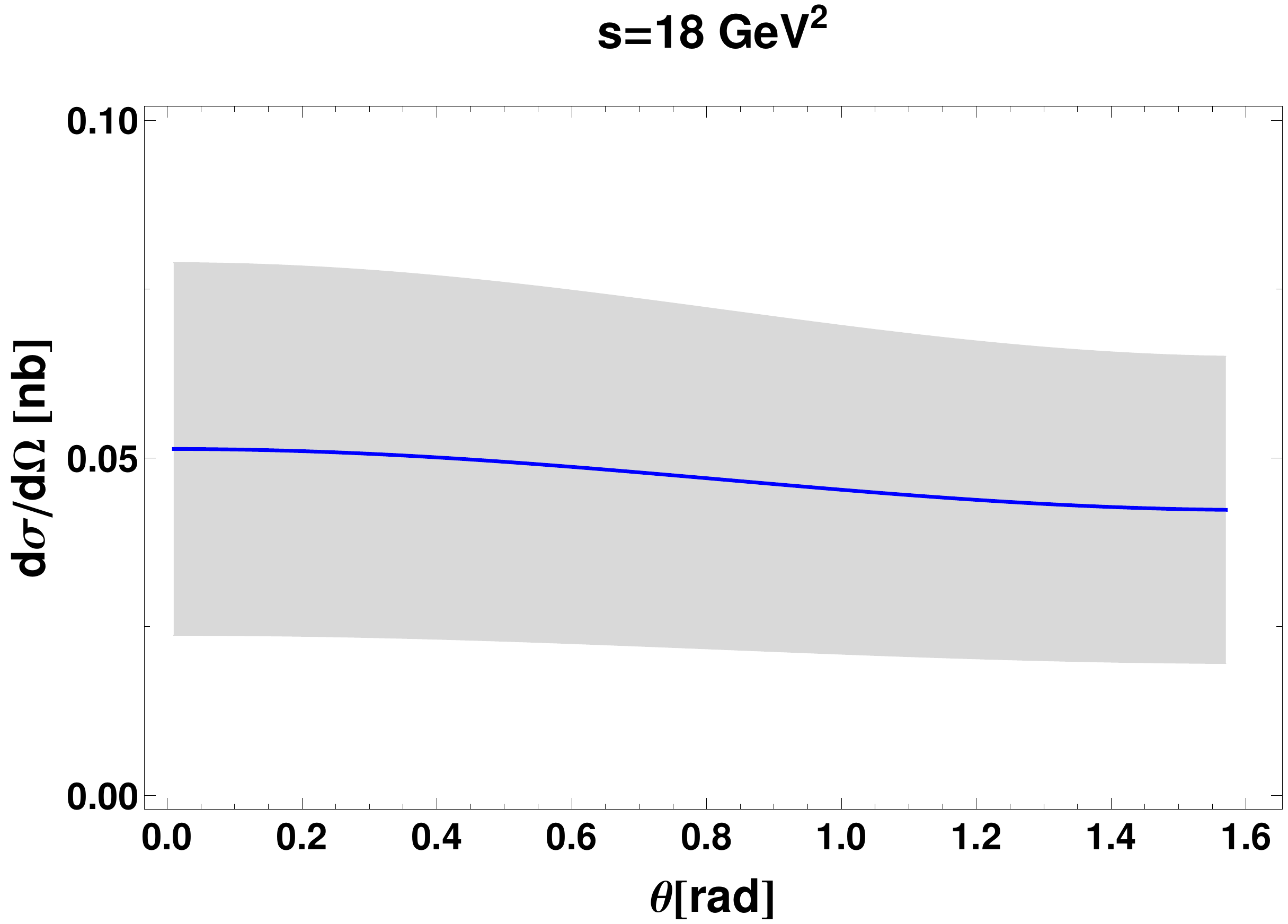}
\hfill
\includegraphics[width=.45\textwidth]{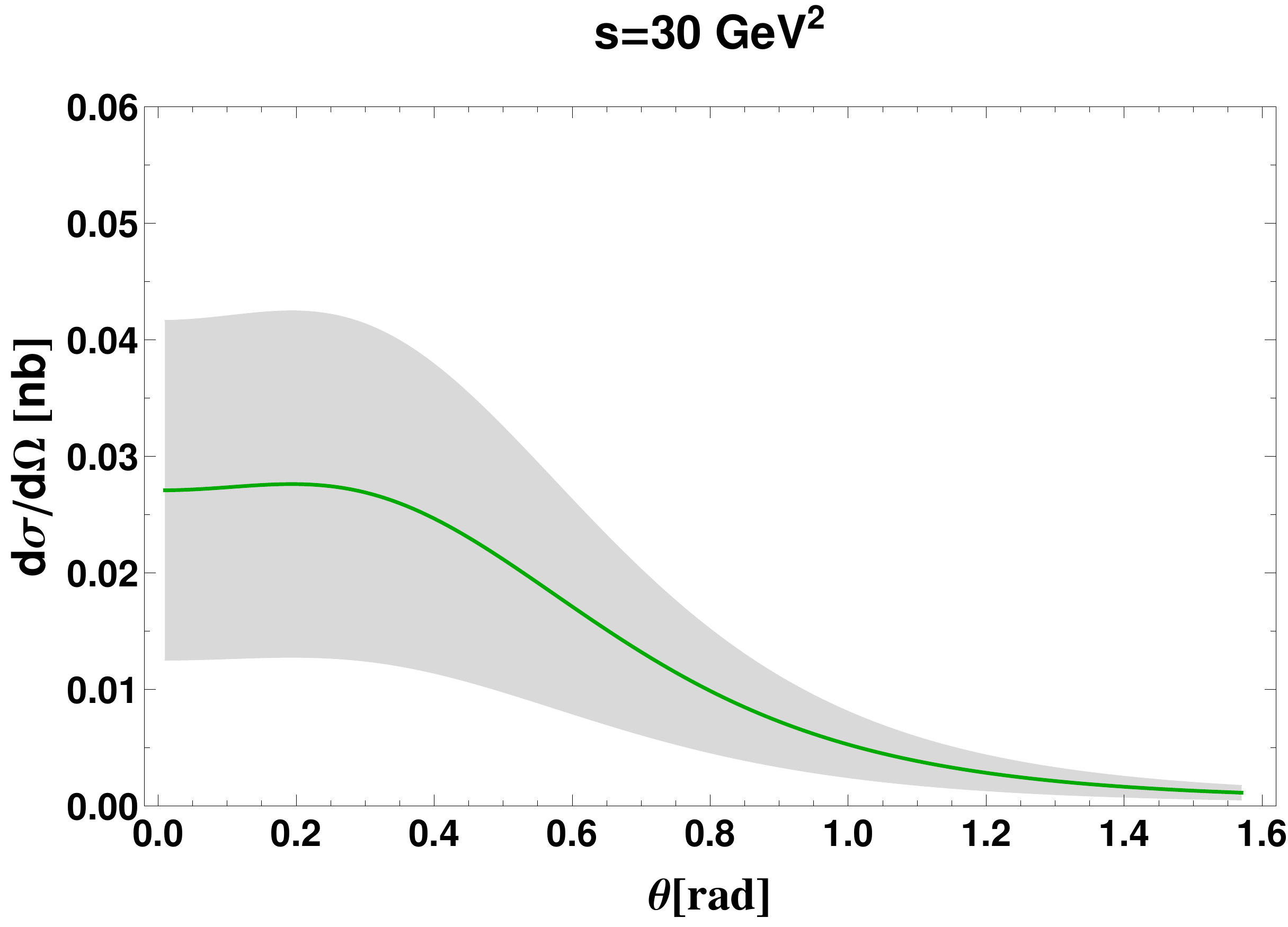}
\caption{The unpolarized \(\process \) differential cross section \( {d \sigma}/{d \Omega} \) versus the CMS scattering angle \( \theta \). The gray error bands show the effect of uncertainties in the LCWFs as discussed in the text.}
\label{fig_diff_cs}
\end{figure}

In Fig.~\ref{fig_diff_cs} we show the unpolarized \(\process \) differential cross section  $d\sigma/d\Omega$ versus the CMS scattering angle \( \theta\) for two energies, one close to the reaction threshold, the other well above. With increasing energy a pronounced peak in forward direction is seen to emerge which can be mainly attributed to the behaviour of the $p\to\Lambda_c^+$ GPDs.
The gray bands take the uncertainties of the (heavy hadron) LCWFs into account. For \( a_\Lambda \) we assume an error of \( \pm 10 \% \), the valence Fock state probability of the \( \mylambda\) is varied in the range \(0.7 - 1 \) and the valence Fock state probability of the D-meson is varied in the range \(0.8 - 1 \).

\begin{figure}[b!]
\centering
\includegraphics[width=.45\textwidth]{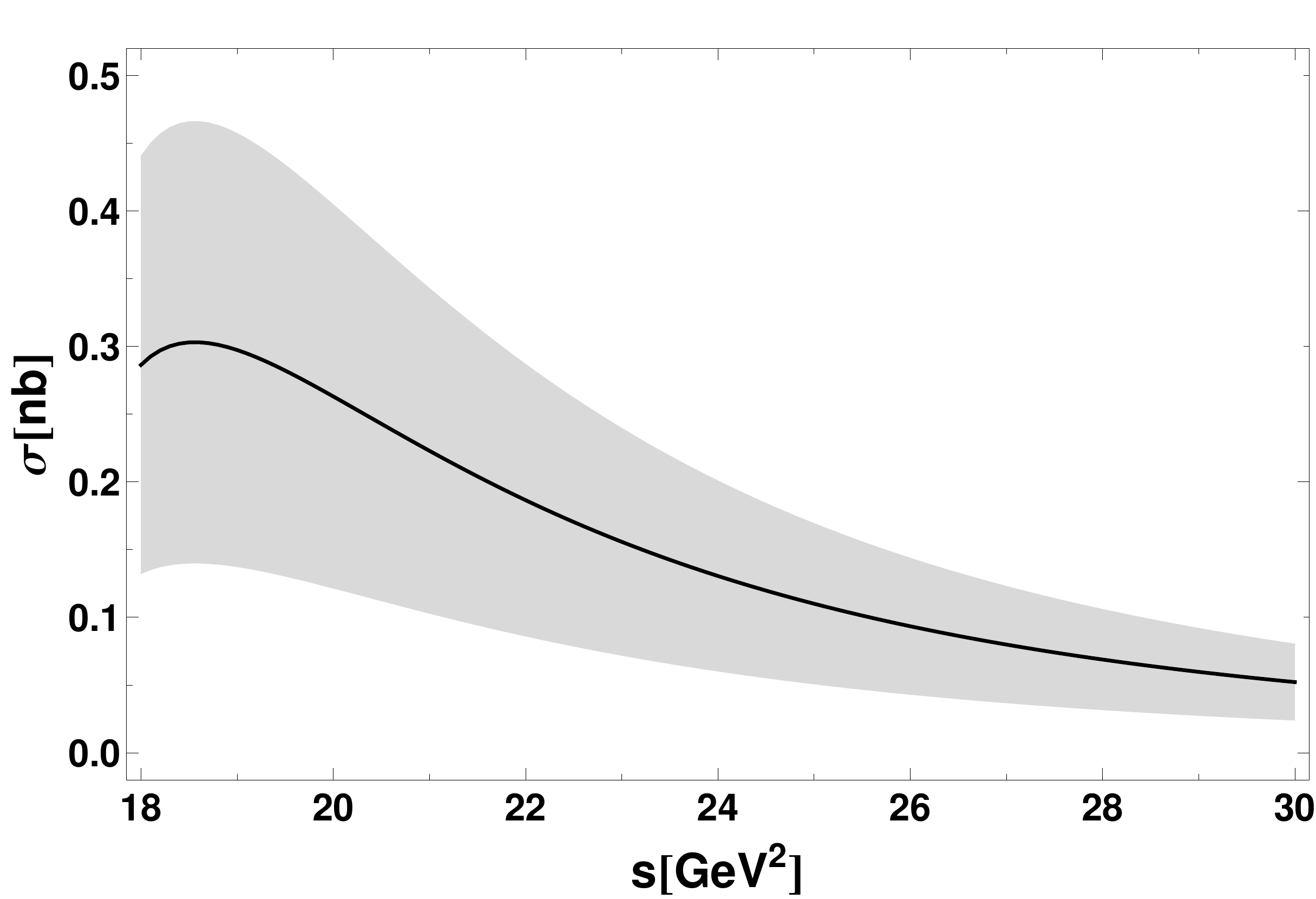}

\caption{The integrated \(\process \) cross section \( \sigma \) versus Mandelstam \( s \).}
\label{fig_int_cs}
\end{figure}

In Fig.~\ref{fig_int_cs} we show the integrated cross section versus Mandelstam \( s \).
The absolute size of the integrated cross section is less than \( 1~\mbox{nb}\) which is about two to three orders of magnitude smaller than the prediction of Refs. \cite{Rekalo1,Rekalo2}, where a hadronic interaction model, based on the ``effective Lagrangian method'', was used.
It is also more than one order of magnitude smaller than the one for \( p \bar{p} \rightarrow \overline{D^{0}} D^{0}\) found in Ref. \cite{schweig2}.
The big discrepancy between our approach and the hadronic interaction model, reveals the necessity of experimental data to discriminate between the different dynamical models.
This experiments could also be helpful to determine the role of the intrinsic charm-quark content in the proton sea. The experimental finding of a much larger cross section could only be explained within our approach if the charm-quark content of the proton sea was indeed non-negligible.
That would require an extension of our approach to account for contributions in which a $\bar{c}$-quark of the nucleon sea goes into the $\overline{D^0}$.
Such contributions are associated with new $p\rightarrow \Lambda_c^+$ transition GPDs which describe the emission of a $\bar{c}$-quark from the proton and the reabsorption of a $\bar{u}$-quark by the remnant of the proton. Experiments on $\process$ could thus provide valuable information on the charm content of the nucleon sea.

\section{Acknowledgements}
We want to thank the organizers of the ``Photon 2013'' conference.
ATG and SK are supported by the Fonds zur F\"orderung der wissenschaftlichen Forschung in \"Osterreich via Grand No. J3163-N16 and FWF DK W1203-N16, respectively.


\begin{thebibliography}{99}

\bibitem{Kofler:2013} Kofler S., Diploma thesis, University of Graz (2013).

\bibitem{Diehl:1998}
Diehl M., Feldmann T., Jakob R. and Kroll P.,
Eur.\ Phys.\ J  {\bf{C8}}, 409 (1999).

\bibitem{Huang:2000}
Huang H. W. and Kroll P.,
Eur.\ Phys.\ J  {\bf{C17}}, 423 (2000).

\bibitem{schweig1}
Goritschnig A.T., Kroll P. and Schweiger W.,
Eur.\ Phys.\ J  {\bf{A42}}, 43 (2009).

\bibitem{schweig2}
Goritschnig A.T., Pire B. and Schweiger W.,
Phys.\ Rev.\ D  {\bf{87}}, 01407 (2013).

\bibitem{Diehl:2000xz}
  M.~Diehl, T.~Feldmann, R.~Jakob and P.~Kroll,
  Nucl.\ Phys.\ B {\bf 596} (2001) 33
   [Erratum-ibid.\ B {\bf 605} (2001) 647]
  [hep-ph/0009255].

\bibitem{Bolz:1996sw}
Bolz J. and Kroll P.,
Z.\ Phys.\ A {\bf{356}}, 327 (1996).

\bibitem{Korner:1992uw}
K$\text{\"o}$rner  J.G. and Kroll P.,
Phys.\ Lett.\ B {\bf{293}}, 201 (1992);
Z.\ Phys.\ C {\bf{57}}, 383 (1993).

\bibitem{Isgur}
Isgur N. and Wise  M. B.,
Nucl.\ Phys.\ B {\bf{348}} (1991).

\bibitem{PDG}
Beringer J. and et al.,
Phys.\ Rev.\ D {\bf{86}}, 01001 (2012).


\bibitem{Rekalo1}
Rekalo M. P. and Tomasi-Gustafsson E.,
Phys.\ Lett.\ B {\bf{500}}, 53 (2001).

\bibitem{Rekalo2}
Rekalo M. P. and Tomasi-Gustafsson E.,
Phys.\ Rev. \ D {\bf{69}}, 094015 (2004).


\end{thebibliography}
\end{document}